\documentclass[aps,prl,showpacs,reprint,groupedaddress]{revtex4-1}
\usepackage{color}
\usepackage{setspace}
\usepackage{verbatim}
\usepackage{graphicx}
\usepackage{subfigure}
\usepackage{amsmath}
\usepackage{soul}
\usepackage{multirow}

\definecolor{orange}{rgb}{1.0,0.76,0.02}

\setstcolor{blue}

\begin{document}

\title{Template induced precursor formation in heterogeneous nucleation - Controlling polymorph selection and nucleation efficiency}

\author{Grisell D\'{i}az Leines}
\email{gd466@cam.ac.uk}
\affiliation{Yusuf Hamied Department of Chemistry, University of Cambridge, Cambridgeshire CB2 1EW, United Kingdom}
\author{Jutta Rogal}
\affiliation{Department of Chemistry, New York University, New York, NY 10003, USA}
\affiliation{Fachbereich Physik, Freie Universit{\"a}t Berlin, 14195 Berlin, Germany }

\begin{abstract}

We present an atomistic study of heterogeneous nucleation in Ni employing transition path sampling, which reveals a template precursor-mediated mechanism of crystallization. Most notably, we find that the ability of tiny templates to  modify the structural features of the liquid and promote the formation of precursor regions with enhanced bond-orientational order is key to determine their nucleation efficiency and the polymorphs that crystallize. Our results reveal an intrinsic link between structural liquid heterogeneity and the nucleating ability of templates, which significantly advances our understanding towards the control of nucleation efficiency and polymorph selection.

\end{abstract}
%\pacs{82.20.Db,82.20.Kh,05.10.Gg,05.10.-a}
%80.20.Db-Transition state theory and statistical theories of rate constants
%05.10.Gg 	Stochastic analysis methods (Fokker-Planck, Langevin, etc.)
%05.10.-a 	Computational methods in statistical physics and nonlinear dynamics 
%82.20.Kh 	Potential energy surfaces for chemical reactions 
\maketitle

%----------------------------------------------Introduction-------------------------------------------------------------

%\section{Introduction}

Gaining control on the emerging polymorphs during 
crystal nucleation is of crucial importance for the synthesis and design of nanomaterials with targeted properties. The presence of impurities and interfaces significantly modifies induction times and the selection of preferred polymorphs during crystallization, which has fuelled extensive research  that focuses on understanding what determines the nucleating ability of a template.~\cite{Sosso2016,Pouget1455, Jungblut2013,Cacciuto2004, Chen2021,C9CE00404A, Egelhaaf2015,refId0,Fitzner2020} 
Yet, unraveling the kinetic pathways of crystal nucleation
at the nanoscale poses a major challenge~\cite{Blow2021} 
as many systems exhibit complex transitions, and often
nontrivial microscopic correlations between liquid-surface
interactions and template morphologies are observed,
leaving the problem largely unsolved.
Fundamental knowledge 
on template-driven nucleation mechanisms is of key importance 
to shed light on predictive rules that allow 
for better control of crystallization processes.

Longstanding views on heterogeneous nucleation generally propose
that the presence of a template influences the nucleation mechanism to a degree in which the template is commensurate with the growing solid cluster 
%that emerges within the melt, 
in both density and symmetry.~\cite{Turnbull1952} 
However, it is well-documented that a small lattice mismatch, although an important factor, is often not the sole requirement for an efficient and successful crystal template. 
Other factors, like template morphology, absorption, and the local ordering of the contact liquid layer can also largely impact the nucleation mechanisms.~\cite{Fitzner2020, PhysRevLett.108.025502,Jungblut2013,Page2009,doi:10.1063/1.4961652} Yet, our current fundamental knowledge on the key factors that determine the nucleating ability of a template is far from conclusive and remains unsatisfactory to date.

Recent evidence demonstrating the formation of crystalline precursors in
the liquid that promote the nucleation of pre-selected polymorphs~\cite{Gebauer2014,tenWolde1997,tenWolde1999,Zhang2007,PhysRevLett.108.225701,Russo2012} has raised great interest in understanding the impact of structural and dynamical heterogeneity in the supercooled liquid on the crystallization mechanism. 
%and its possible link to the nucleating ability of a templates.
%In light of these findings, the nucleating ability of a template is bound to be linked to modifications of the local structural and dynamical features of the liquid.
Precursor-induced crystallization processes, often referred to as two-step nucleation mechanisms, are characterized by the initial formation of pre-ordered regions in the liquid that exhibit changes in bond-orientational order, density, or mobility, and facilitate the formation of  crystal nuclei, presumably by decreasing the crystal-liquid interfacial free energy.~\cite{DLeines2018,Tanaka2012, PhysRevX.8.021040}
Russo {\it et al.}~\cite{Russo2012} showed that pre-ordered regions in hard spheres liquids act as polymorph precursors by pre-selecting the polyhedra with the closest symmetry to those of the crystalline phases that nucleate. Previously, we have shown that pre-ordered liquid regions also act as preferred nucleation sites that pre-determine the polymorphs in metallic systems with 
%different bulk structures, like 
face-centred cubic (fcc)~\cite{DLeines2017,DLeines2018} and body-centred cubic (bcc)~\cite{doi:10.1063/5.0017575} bulk structures.
Several studies on nucleation of ice,~\cite{Fitzner2009} metals,~\cite{DLeines2017,DLeines2018,Zhang2019} 
hard spheres,~\cite{PhysRevLett.105.025701,PhysRevLett.96.175701} and colloidal models~\cite{Lechner2011a,Tan2014} 
have corroborated an existing link between liquid heterogeneity and the enhancement of the nucleation probability, as well as the selection of polymorphs  during the first stages of crystallization. 
Indeed, a recent work on model binary liquids with tunable glass-forming ability~\cite{PhysRevX.8.021040} showed that the structural differences within the supercooled liquids are key to control the glass  
and crystal-forming 
ability. 
%The increase in structural similarity between liquid and crystal, quantified by an interface energy penalty parameter, was found to promote the formation of crystal precursors that increase the crystallization ability of a system. 
%that pre-select particular polymorphs
%When the melt 
%exhibits more structural similarity to the crystals , these polymorph 
%precursors will emerge and enhance the crystal nucleation by reducing 
%the liquid-crystal interfacial free energy penalty~\cite{XXX}. 
Fundamental knowledge of the relation between structural and dynamical heterogeneity of supercooled liquids and  nucleation is therefore opening up a new perspective in our understanding of crystallization mechanisms and providing novel possibilities to control polymorph outcomes.
Yet, it is largely unexplored how nucleating agents and interfaces impact the structural and dynamical characteristics of the supercooled liquid, and how this is connected with heterogeneous nucleation mechanisms.

In this work, we tackle the above question by investigating how small seeds with different crystal structures modify the structural characteristics of the supercooled liquid and ultimately impact the crystallization mechanism.The importance of studying small seeds to improve our understanding of crystallization mechanisms and towards the selective control of materials properties, has been evidenced in~\cite{Cacciuto2004,Villeneuve_2005,Jungblut2013, Jungblut_2011, Jungblut2016b,PhysRevLett.100.108302}.  Experimentally, this can be achieved with optical tweezers~\cite{doi:10.1063/1.1522397,van_blaaderen_template-directed_1997,doi:10.1063/1.1784559,C9SM01297D,C0SM01219J} where templates are created by fixing the positions of atoms in the liquid to investigate the impact on the polymorphs that crystallize and the rates. Our study focuses on  Ni, for which we have previously identified 
strong spacial-temporal correlations between pre-ordered liquid regions and the nucleation process.~\cite{DLeines2017,DLeines2018} 
The role of liquid pre-ordering during template-driven crystallisation in Ni is, however, unknown.
To enable an efficient sampling of the nucleation process, 
%which constitutes a {\it rare event} taking place on extended timescales, 
we employ transition interface sampling (TIS).~\cite{Dellago2002,VanErp2005,VanErp2007}
In TIS, an unbiased ensemble of all possible nucleation pathways between the solid and liquid state~\cite{VanErp2005,VanErp2007,Rogal2010} is computed. Kinetic and dynamical properties, such as  nucleation barriers and  rate constants, can be obtained by reweighting each path in the ensembles according to its correct probability.~\cite{Rogal2010} 
%yielding the reweighted path ensemble (RPE).~\cite{Rogal2010} 
Our analysis of  heterogeneous nucleation pathways reveals a novel template-driven mechanism, where the ability of the seeds to enhance the nucleation probability of selected polymorphs is not directly determined by the degree of lattice mismatch between the seeds and the crystalline bulk phase, but by their ability to promote the formation of precursor regions  that modify the nucleation probability and facilitate the emergence of specific polymorphs.

\begin{figure}[tb]
\includegraphics[width=9.0cm,clip=true]{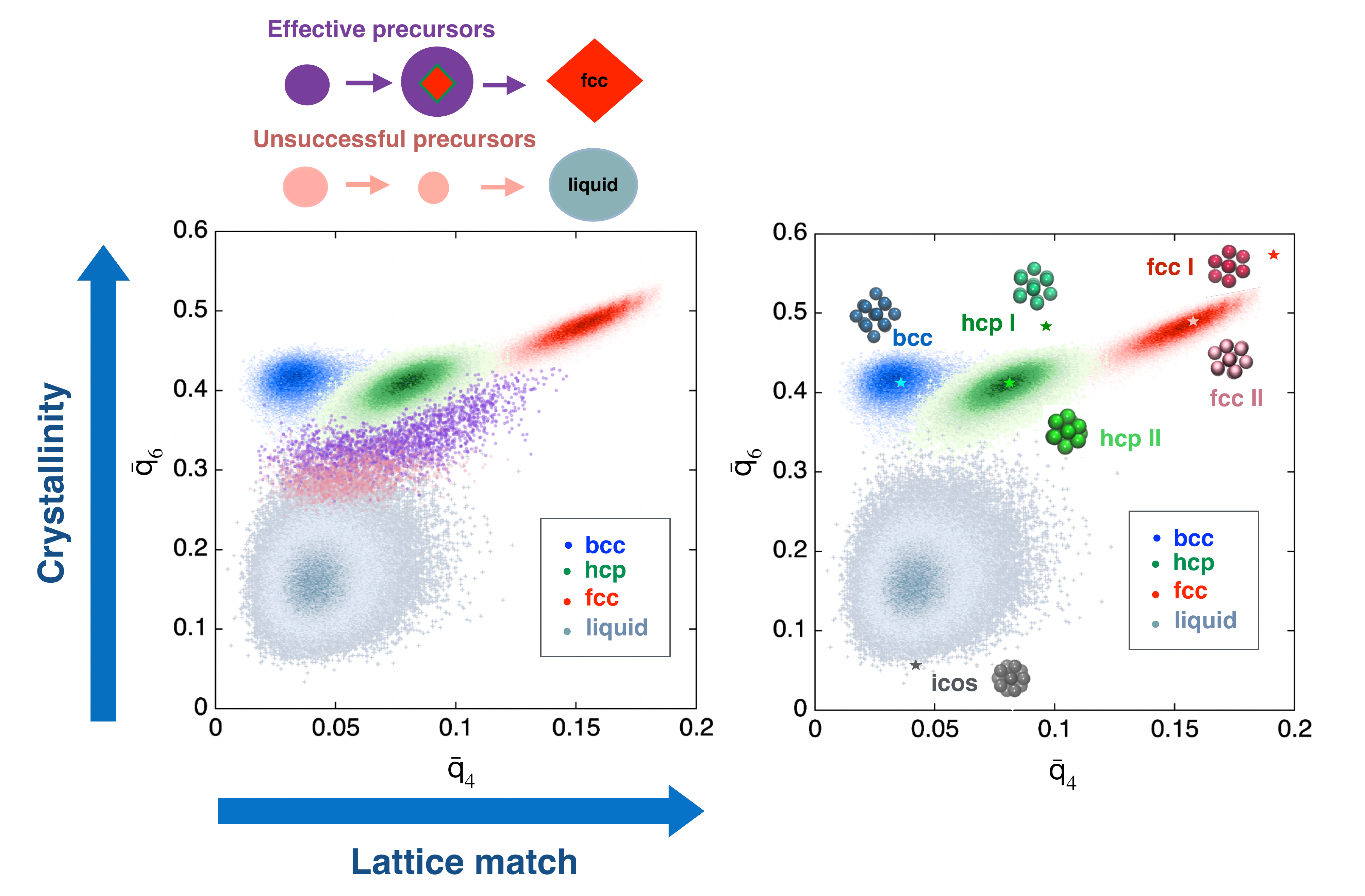}
\caption{\label{fig1}  (Left) $\bar{q}_4-\bar{q}_6$ values for pre-order liquid particles in pre-critical clusters ($n_s=50$) of the TPE that  reach critical size and crystallize  (purple dots), and  in the pre-critical clusters of smaller sizes that commit back to the liquid phase (pink dots). (Left top) Schematic representation of pre-ordered liquid clusters with higher bond-orientational order (purple) that crystallize and lower bond-orientational order (pink) that melt. (Right)  $\bar{q}_4-\bar{q}_6$ values of selected seeds with different structures. 
}
\end{figure}

All simulations were performed in the NPT ensemble employing an embedded atom method (EAM) potential for Ni~\cite{Foiles86} and the \textsc{lammps} code~\cite{Plimpton1995} as molecular dynamics (MD) driver together with a python wrapper for the TIS simulations (computational details in Supplemental Material (SM)~\cite{supplemental}). 
%As order parameter 
To discriminate between solid-like and liquid-like particles, the approach 
%based on Steinhardt parameters 
introduced by ten Wolde and Frenkel~\cite{Steinhardt1983,Auer2005} was used (details in SM~\cite{supplemental}). 
The size of the largest solid cluster, $n_s$, is determined via a clustering algorithm.
For the local identification of crystal structures such as fcc, hexagonal close-packed (hcp), bcc, as well as liquid and pre-structured liquid, we use 
%a combination of 
%the 
averaged Steinhardt parameters~\cite{Lechner2008} $\bar{q}_4,\bar{q}_6$.
{\it Pre-structured liquid} particles are 
particles that exhibit higher bond-orientational order than the liquid, but less than any of the crystalline phases and fall outside the corresponding regions on the $\bar{q}_4,\bar{q}_6$ map (Fig.~\ref{fig1}). 
Details can be found in Ref.~\onlinecite{DLeines2017} and in SM~\cite{supplemental}. 

In order to associate the nucleating ability of  templates with the formation of efficient  precursors in the melt, we first investigate 
the structural characteristics of pre-ordered regions in the liquid that act as preferential sites for homogeneous  nucleation in Ni.
To this end, we analyse pre-critical clusters obtained from 400  trajectories  of the transition path ensemble (TPE) of homogeneous nucleation in Ni~\cite{DLeines2017,DLeines2018} at $\Delta T/ T_m=20\%$ undercooling. 
In the crystallization mechanism found in pure Ni,~\cite{DLeines2017} 
pre-critical clusters with $n_s\leq 50$ are mostly composed of pre-structured liquid ($> 90\%$)
and facilitate the subsequent nucleation of crystallites within the core of these precursor regions.~\cite{DLeines2017}
%Pre-critical clusters are mostly composed  of pre-structured liquid ($> 90\%$).
Thus for our analysis of precursors, we harvest pre-critical clusters with $n_s=50$ from 400  trajectories of the RPE. 
The left graph in Fig.~\ref{fig1} shows a reference $\bar{q}_4,\bar{q}_6$ map of the crystalline phases and the liquid together with  the distribution of $\bar{q}_4,\bar{q}_6$ values of  pre-structured liquid particles in these pre-critical clusters from trajectories that successfully nucleate into the crystal phase (purple), in comparison to clusters 
that become unstable and dissolve (pink). 
The pre-structured liquid clusters that continue to grow beyond the critical size exhibit a clear increase in bond-orientational order.
Furthermore,
%the $\bar{q}_4,\bar{q}_6$ values of  
pre-ordered clusters that serve as nucleation sites tend to contain hcp/fcc-like structural motifs, resembling the bulk structure that crystallizes. In contrast, pre-critical clusters that dissolve are characterized by lower bond-orientational order with structural features closer to the liquid. Therefore, {\it effective} precursor regions with higher bond-orientational order and hcp/fcc-like features in the liquid template crystal nucleation by providing preferential sites for critical fluctuations. 

Having established that fluctuations in the liquid with higher $\bar{q}_4,\bar{q}_6$ values promote crystallization during homogeneous nucleation, we investigate how small Ni seeds with different crystal structures modify the structural features of the liquid and the crystal precursors, in connection with their nucleating ability. We select seeds with fcc, bcc, hcp, and icosahedral structures and various degrees of crystallinity, 
%or bond-orientational order, 
as shown in the $\bar{q}_4,\bar{q}_6$ map in Fig.~\ref{fig1} (right). The tiny seeds consist of atoms within the first coordination polyhedron.
We include two types of fcc and hcp seeds, labelled as fcc/hcp I and fcc/hcp II, respectively, which have the same symmetry and average bond length  but differ in crystallinity. Fcc/hcp~I correspond to perfect polyhedra, while fcc/hcp~II  represent polyhedra with  thermal distortions. 
%chosen at the maximum of the $\bar{q}_4,\bar{q}_6$ reference distributions (Fig.~\ref{fig1}) 
%chosen to have $\bar{q}_4,\bar{q}_6$ values at the maximum  of the reference distributions for bulk fcc and hcp  at $\Delta T/T_m = 20 \%$.
The seeds are inserted in the liquid and have a fixed position. 
To test the approximation of fixed seeds, we have also performed simulations with an fcc seed that is allowed to vibrate, yielding comparable results (see SM~\cite{supplemental}).

\begin{figure}[tb]
\includegraphics[width=9.2cm,clip=true]{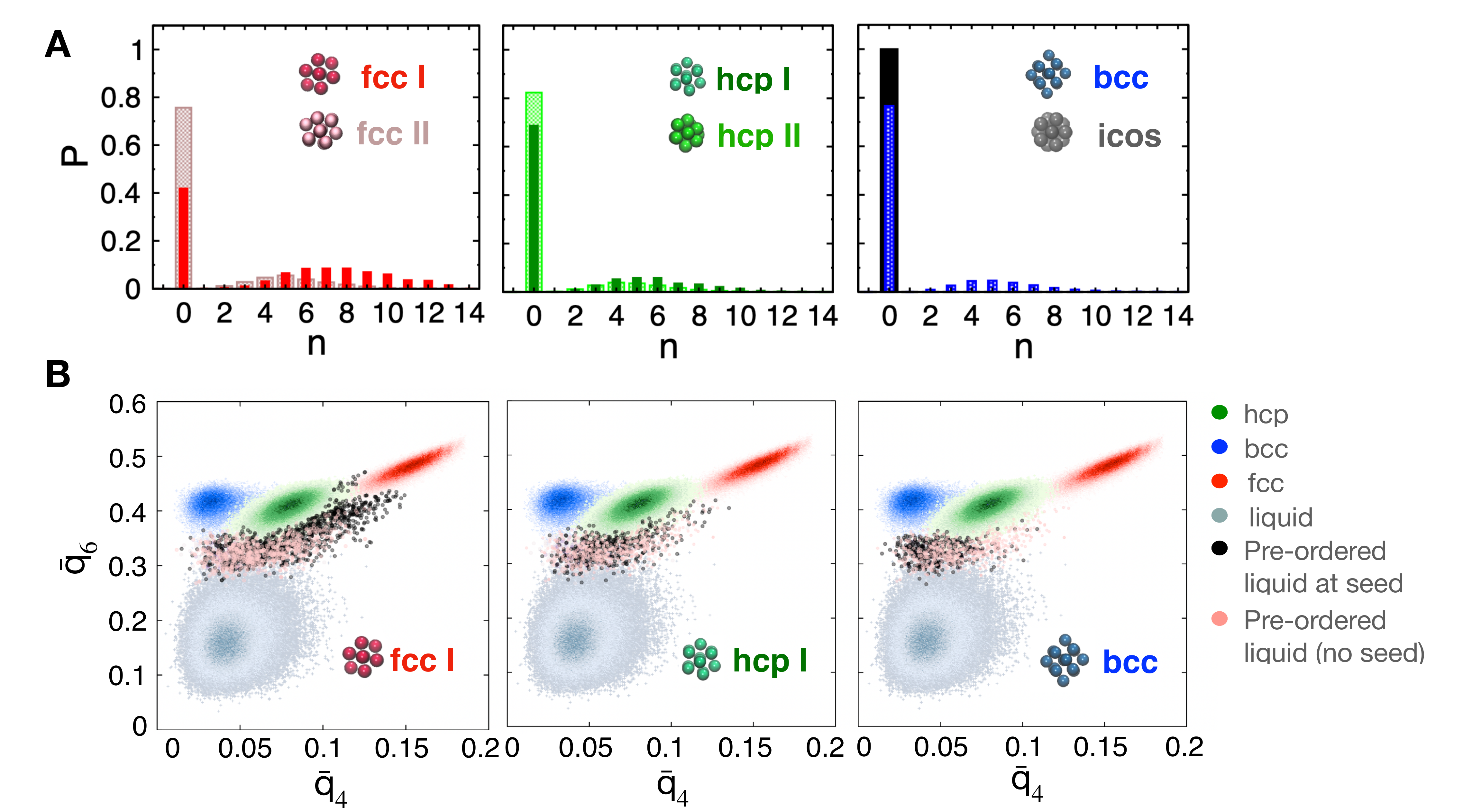}
\caption{\label{fig2}  A) Frequency of  the formation of pre-ordered clusters at the seed in supercooled liquid. Shown is the number of atoms $n$ of the seed that belong to the largest pre-ordered cluster that emerges during fluctuations in the supercooled liquid, at $\Delta T/T_m=0.20$. 
%Histograms for the fcc I (red), fcc II (pink), hcp I (green), hcp II (light green), bcc (blue) and icosahedral (gray) seeds are shown. 
%The arrows indicate the positions of the maximum number of seed atoms. 
%that belong to the largest pre-ordered liquid clusters. 
B)  $\bar{q}_4-\bar{q}_6$ values for  pre-ordered liquid particles that belong to the largest pre-critical cluster that emerges within homogeneous liquid (pink) and at selected seeds (black) in the liquid. 
Additional distributions for all other seeds are shown in the SM~\cite{supplemental}.
%The distributions are shown for fcc I, hcp I and bcc seeds, to illustrate representative cases (all seeds shown in Supp. Info.)
}
\end{figure}

We first characterize the ability of the seeds to promote the formation of pre-ordered regions in the liquid  that emerge from pre-critical fluctuations (typically $n_s<30$), by considering two aspects: ({\it i}) the frequency of formation of the largest pre-structured liquid cluster at the seed, and ({\it ii}) the degree of bond-orientational order and the structural hallmarks of the pre-structured regions formed at the seed. To address these two aspects, we perform five independent MD simulations of liquid  Ni over 5~ns at $\Delta T/T_m=0.20$ for each seed. The distribution of the number of seed atoms that belong to the largest pre-structured liquid cluster is shown in Fig.~\ref{fig2}A, the $\bar{q}_4,\bar{q}_6$ values of the pre-ordered clusters that emerge at the seed in  Fig.~\ref{fig2}B (results for other seeds  shown in SM~\cite{supplemental}).
If none of the  seed atoms are included in the pre-structured cluster ($n=0$), pre-critical fluctuations occur far from the seed. 
This is further supported by evaluating the distribution of minimum distances between the seed and the largest pre-structured cluster (see SM~\cite{supplemental}).
Our results show that the most commensurate seeds with larger bond-orientational order (fcc~I) clearly promote more frequent pre-critical fluctuations of pre-ordered  clusters at the seed (Fig.~\ref{fig2}A) and enhance the crystallinity and fcc-like symmetries in the pre-structured liquid, in comparison to pre-odered regions found in homogeneous liquid  (Fig.~\ref{fig2}B). Indeed, for the fcc~I seed, ~60\% of the pre-ordered clusters form at the seed, while for hcp~I and fcc~II seeds, with lower crystallinity, ~34\% and ~24\% of the pre-ordered clusters form at the seed, respectively.
For hcp II and bcc seeds, which are expected to be less efficient nucleating agents, only ~12\% and ~18\% of the pre-ordered clusters emerge at the seed, respectively, while
pre-critical fluctuations always occur far from the icosahedral seed, indicating that pre-ordering in the liquid is inhibited in its vicinity. 
%Moreover,  the maximum in the distributions for $n >0$ in Fig.~\ref{fig2}A  indicates that more seed atoms are included in the pre-ordered clusters depending on their ability to promote pre-critical fluctuations at the seed. 
%and thus it inhibits the formation of precursors in the neighbor region. 
Overall, we observe that seeds with larger crystallinity  promote an increase in crystallinity and fcc-like structural features in the precursors, evidenced by the shift in the distributions in Fig.~\ref{fig2}B. In contrast, seeds with lower crystallinity (bcc and hcp II) promote hcp and bcc-like pre-ordering in the liquid with negligible increase in bond-orientational order. It is also interesting to note that, although fcc~I, fcc~II, and the vibrating fcc seed (see SM~\cite{supplemental}), as well as hcp~I and~ II share a common crystal structure, their variation in crystallinity results in significant differences in the frequency of formation of precursors at the seed and the structural features of these regions. 
Indeed, seeds with the potential to promote  frequent formation of pre-ordered clusters with high crystallinity and fcc-like order in the liquid are bound to enhance the 
%frequency of 
formation of {\it effective}  precursors, i.e. pre-ordered regions which become active sites for nucleation,
and, presumably, such seeds exhibit a larger nucleating ability.
%(as shown above for homogeneous nucleation) and thus enhance the nucleation probability in the melt. 

\begin{figure*}[tb]
\includegraphics[width=18.5cm,clip=true]{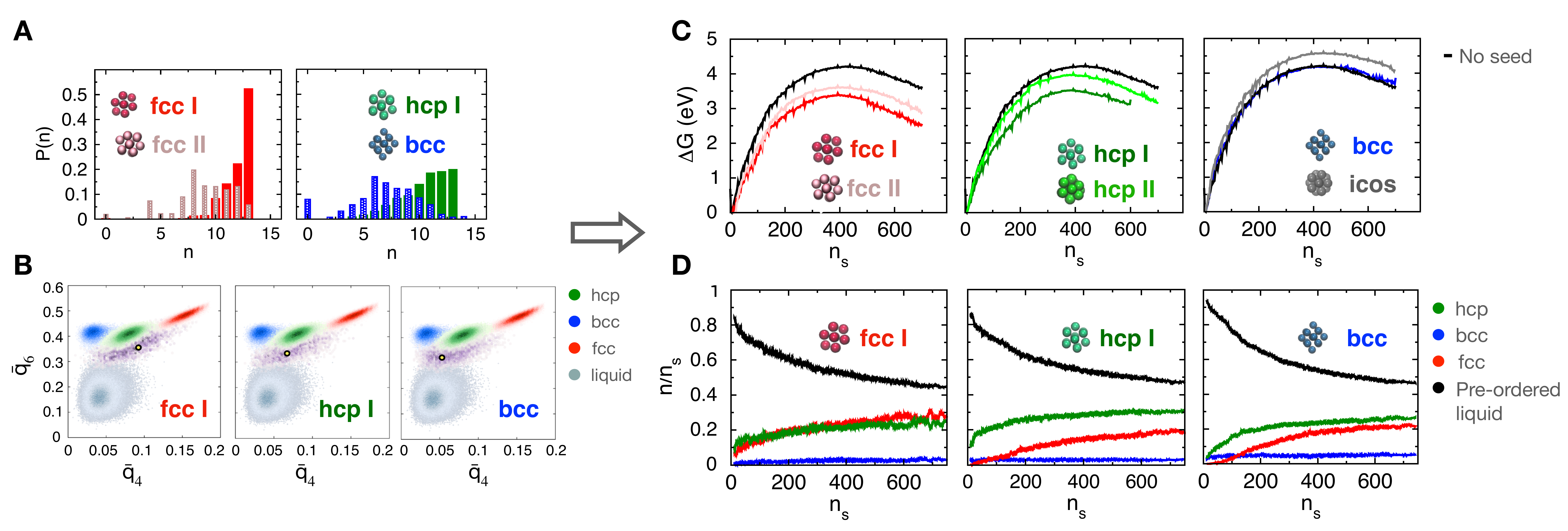}
\caption{\label{fig3} (A) Distributions of the number of seed atoms $n$  included in pre-critical clusters ($n_s=50$) that successfully grow and crystallize (effective precursors). The pre-critical configurations were obtained from the TPE. (B) $\bar{q}_4,\bar{q_6}$ distributions of the effective precursors in the presence of different seeds. The yellow circle indicates the maximum of the distribution. (C) Free energy profiles  $\Delta G (n_s)$ and nucleation barriers for crystallization in the presence of various seeds, in comparison to homogeneous nucleation in Ni. (D) Structural composition of the growing nucleus in the presence of various seeds. }
\end{figure*}

In order to establish the connection between structural changes in the liquid induced by the seeds and the nucleation mechanisms, we performed TIS simulations for all seeds 
%obtaining at least 300 decorrelated paths per interface with a total of 1500 MC moves for each ensemble
(computational details in SM~\cite{supplemental}).  In all cases, the analysis of the structural compositions of the growing nuclei in the TPE (see SM~\cite{supplemental}) reveals a precursor-induced crystallization mechanism, similar to the one found in homogeneous nucleation~\cite{DLeines2017}, where we find a prestructured region of long-life that is present before the crystal clusters nucleate. An analysis of the spatial location of the seeds within solid clusters of pre-critical and critical sizes (see SM~\cite{supplemental}) 
%obtained by computing the minimum distance of the seeds to the nucleus surface, 
shows that if a seed is part of the growing nucleus, a pre-ordered liquid cluster  initially forms at the seed and grows in its surroundings. As the cluster reaches critical size, the seed remains predominantly located at the surface of the nucleus, where it is surrounded mostly by pre-structured liquid atoms and random-hcp (see movie of nucleation trajectory in SM~\cite{supplemental} ). These findings strongly suggests that the seeds impact the nucleation probability and mechanism 
%to a degree in which they promote  
by promoting the initial formation of precursors in the supercooled liquid instead of directly templating the formation of the crystalline phase. Indeed, if the nucleation events at the seeds were driven solely by the initial enhancement of translational order, we would rather expect fcc crystallites to nucleate right at the seeds without an intermediate precursor and, consequently, the seeds to be located in the core of the critical clusters. Therefore, the overall efficiency of a seed to promote nucleation events and, in general, to modify the nucleation mechanism can be characterized by: %three criteria:  
%1) the ability of the seeds to promote heterogeneous nucleation,
({\it i}) the ability to promote the formation of {\it effective} precursors, 
%i.e. pre-ordered liquid regions with enhanced crystallinity and fcc-like structural hallmarks that yield critical fluctuations, 
({\it ii}) the potential to reduce the free energy barrier, and ({\it iii}) the propensity to template the formation of favorable polymorphs.

We asses these three aspects for all seeds by analyzing  configurations from at least 500 liquid-solid pathways in the TPE.
%An analysis of the distributions of the number of seed atoms $n$ included in pre-ordered clusters of pre-critical size ($n_s=50$) that successfully grow and crystallize ({\it effective} precursors) shows that seeds with the highest crystallinity and commensurability, such as fcc~I and~II, are part of the growing clusters for $~99$\% of the solid-liquid pathways and thus successfully promote nucleation events at the seed (see SM~\cite{supplemental}).
Fig.~\ref{fig3}A shows the distributions of the number of seed atoms $n$ included in {\it effective} precursors of pre-critical size ($n_s=50$), that is  pre-ordered clusters that successfully grow and crystallize. Seeds with the highest crystallinity and commensurability, such as fcc~I and~II, are part of the growing clusters for $~99$\% of the solid-liquid pathways and thus successfully promote nucleation events at the seed.   
For other seeds with lower crystallinity and commensurability (hcp~I, hcp~II, bcc), there is a stronger competition between heterogeneous and homogeneous nucleation pathways. In case of hcp~I seeds, these two pathways appear to follow separate channels which results in a strong dependency of the TPE on the initial trajectory. If the hcp~I seed is initially part of the crystal cluster, the TPE is predominantly composed of heterogeneous nucleation pathways 
while initial paths with the hcp I seed far from the growing cluster result in a TPE with predominantly homogeneous nucleation pathways. The overall nucleation mechanism in the presence of hcp~I seeds would be given by a properly weighted average of the two pathways. %For bcc and hcp seeds, nucleation events can take place at the seed ($n>5$), as well as in the bulk liquid ($n=0$).
Icosahedral seeds are never identified as part of the growing cluster, implying that nucleation occurs mostly far away from the seeds.
Interestingly, the distributions are noticeably different for all the seeds, even for those with shared crystal structures: while seeds with high crystallinity, such as fcc~I, are mostly included in the clusters ($n>10$), partial attachment of the crystalline clusters to other seeds with lower crystallinity (fcc~II, hcp II, bcc) is more frequently observed ($n<6$), implying significant variations in the ability of the seeds to promote the formation of effective precursors.
These results are consistent with our findings for pre-critical fluctuations shown in Fig.~\ref{fig2}A.

The ability to promote {\it effective} precursors is further assessed by analyzing the distributions of $\bar{q}_4,\bar{q}_6$ values of the pre-critical clusters ($n_s=50$) that successfully grow and crystallize at the seeds.  The shift and spread in the distributions of $\bar{q}_4,\bar{q}_6$ values shown in Fig.~\ref{fig3}B clearly illustrates  
that pre-structured liquid atoms in precursors that form at 
seeds with higher crystallinity (fcc~I) display a significant increase in bond-orientational order and fcc-like ordering (marked with a yellow circle in Fig.~\ref{fig3}B). This change in the structural characteristics of the pre-ordered region is likely to reduce induction times and enhance  nucleation rates in comparison to  other seeds and homogeneous nucleation.
Correspondingly, 
the  fcc~I seed reduces the free energy barrier significantly by $\sim 0.85$~eV compared to homogeneous nucleation with $\Delta G_{\text{homo}}^* = 4.21$ eV (Fig.~\ref{fig3}C). The less effective fcc~II and hcp~I seeds, 
with comparable crystallinity, 
%which have comparable crystallinity but are less efficient than fcc~I in enhancing the crystallinity and fcc-like ordering in the precursors, 
reduce the  nucleation barrier extracted from  heterogeneous nucleation pathways by only $\sim 0.6$ and $\sim 0.69$ eV, respectively. 
The hcp II and bcc seeds  enhance bond-orientaional order in the precursors only minimally and, thus, the free energy barriers are 
%not significantly reduced compared 
comparable
to homogeneous nucleation. Interestingly, the nucleation barrier at the icosahedral seed is even increased, implying that this seed acts as an impurity that reduces the nucleation capability in Ni.
Icosahedral seeds inhibit the formation of pre-ordered regions in their surrounding, thus resulting in excluded volume for nucleation sites.
%An explanation for this finding is that the icosahedral seed inhibits the formation of pre-ordered regions in the surrounding regions of the seed and this excluded volume results in a slight reduction of the nucleation probability. 

The impact of the seeds on the formation of different polymorphs is evaluated by analysing the average structural composition of the growing nuclei shown in Fig.~\ref{fig3}D. 
The polymorphs selected during crystal nucleation in Ni correlate strongly with the structural hallmarks promoted by the seeds in the pre-ordered liquid. The fcc~I and~II seeds promote the formation of precursors with enhanced fcc-like hallmarks  and, consequently, yield a rapid and predominant emergence of fcc crystallites within the cores of the crystal precursors and critical nuclei (Fig.~\ref{fig3}D and SM~\cite{supplemental}). In contrast,  for hcp~I, hcp~II, and bcc seeds, which promote the formation of precursors with hcp-like hallmarks, we find a larger fraction of hcp crystallites that compete with fcc. 
%order within the cores of the critical nuclei, bottom graph in Fig.~\ref{fig4}C. 
%(SM~\cite{supplemental}). 
%Our results show that, in addition to lattice mismatch and translational order, the ability of the seeds to template precursors in the liquid, by enhancing the bond orientational order and modifying the structural characteristics of the supercooled liquid, is crucial to enhance the efficiency of the nucleating agents and determine the polymorphs selected during  crystallization in Ni. 

In contrast to the assumptions of classical nucleation theory, where random fluctuations of order within the homogeneous liquid yield crystallization, we have shown  that supercooled liquids exhibit structural heterogeneity that can be  linked directly to crystal nucleation and to the ability of templates to enhance the nucleation probability and modify the polymorphs. We propose a novel heterogeneous nucleation mechanism, where the nucleating ability of tiny seeds and the selection of polymorphs  is not only determined by the lattice match and translational order of the templates, 
%between the crystal phase that nucleates and the template, 
but is strongly linked to the ability of the seeds to promote the formation of precursors in the liquid with enhanced bond-orientational order and favorable structural hallmarks. Previous findings of precursor-mediated crystallization mechanisms in a large variety of systems~\cite{DLeines2017,DLeines2018,Zhang2019,PhysRevLett.105.025701,PhysRevLett.96.175701, Lechner2011a,Tan2014, doi:10.1063/1.4961652, Tanaka2012} suggest that this novel perspective of heterogeneous nucleation could be of relevance for other materials. Our results open new venues to understand and control template-driven crystallization and polymorph selection.

%----------------------------------------------Acknowledgements-------------------------------------------------------------

\section*{Acknowledgements}
We acknowledge financial support by the German Research Foundation (DFG) through project 262052203 and the DFG Heisenberg Programme project 428315600. 
The authors acknowledge computing time by the Center for Interface-Dominated High Performance Materials (ZGH, Ruhr-Universit{\"a}t Bochum, Germany).

%\bibliography{references}
%}
%

\end{document}

\maketitle

\section{MD simulation details}

In all our MD simulations we used an embedded atom method (EAM) potential to model the interactions between Ni atoms~\cite{Foiles86}. For this potential the melting temperature reported is $T_m = 1710$ K~\cite{Hoyt2009}, close to the experimental value
of $T_m = 1728$ K. The MD  package LAMMPS~\cite{Plimpton1995} was used to perform the 
MD simulations.  We use the isothermal-isobaric (NPT) ensemble in all our simulations with a time step of 2~fs
and Nos\'{e}-Hoover thermostat and barostat. All simulations were performed at $P=0$~bar and a temperature $T=1370$~K, 
corresponding to a moderate undercooling of  $20 \%$. The system size is set $N=8788$ atoms, sufficiently large to avoid finite size effects. Minimum image periodic boundary conditions were applied in all  directions for all the simulations. 

The Ni seeds have different structures (fcc, bcc, hcp and icosahedral) and include the atoms within the first shell of nearest neighbors (13 atoms for fcc, hcp and icosahedral seeds  and 15 atoms for bcc). The Ni seeds are inserted in the liquid box and the components of the forces for the seed atoms are set to zero to freeze them in the simulation box. We equilibrate the system for at least 2 ns for each MD simulation. To compare our model of frozen seeds we allow the fcc seed to vibrate by applying a spring force in all directions independently to each atom of the seed, to tether the seed to its initial position. The spring energy is not included in the total potential of the system. 

\section{Transition interface sampling simulation details}

The path sampling simulations were performed using a replica exchange transition interface sampling (RETIS) method~\cite{VanErp2007,Bolhuis2008}. For each seed we perform RETIS simulations with at least
$1500$ moves per interface which included $45\%$ shooting moves~\cite{Dellago1998}, $45\%$ exchange moves~\cite{VanErp2007,Bolhuis2008}, and $10\%$ exchanges between the forward and backward ensembles~\cite{VanErp2007,Bolhuis2008}. Paths were collected in the ensemble after 5 decorrelation steps yielding at least 300 trajectories for each interface, which we previously found to be enough 
for convergence of thermodynamic and kinetic properties in Ni~\cite{DLeines2017}. 
As order parameter for the TIS simulations we use the size of the largest solid cluster, $\lambda = n_s$.
The positions of all interfaces between the stable states are   20, 25, 30,38, 50, 63, 80, 91, 100, 115, 125, 140, 170, 200, 230, 270, 300, 350, 400, 450, 350, 500, 600, 650, 700, adjusted such that there is at least $10\%$ overlap of the corresponding crossing histograms. 
The liquid stable state regions were defined as $n_{s} \leq 20$, while the solid stable states were set to $n_{s} \geq 700$. The solid stable state boundary is set at a value where the crossing probabilities become constant and all trajectories commit to the solid state $B$.

\section{Order parameter and local structure identification}

Sampling nucleation pathways requires an order parameter that can discriminate between solid-like and liquid-like particles. We employ the approach based on Steinhardt order parameters introduced by ten Wolde and Frenkel~\cite{Steinhardt1983, Auer2005}. In this method, a particle shares a solid bond with a neighbor  if  $s_{ij}  =  \sum_{m=-6}^{6} q_{6m}(i)q_{6m}^*(j)> 0.5$, where $q_{6m}$ are the complex vectors defined by spherical harmonics~\cite{Steinhardt1983}. To strengthen the solid definition for particles at the cluster interface, we employ  the average correlation over neighboring atoms $\langle s_{i} \rangle =1/N_\text{nn}\sum s_{ij}$~\cite{Bokeloh2011}. If a particle $i$ has 7 solid bonds and $\langle s_{i}\rangle > 0.6$, it is identified as a solid particle. 
A clustering algorithm then allows us to compute the size of the largest solid cluster $n_s$. 

The local crystal structures of the solid particles are identified by using the averaged  Steinhardt  bond order parameters $\bar{q}_4$ and $\bar{q}_6$~\cite{Lechner2008}. Reference distributions of $\bar{q}_4, \bar{q}_6$ values for crystal structures  were obtained from 10~ns MD simulations of bulk crystals hcp, bcc, fcc and liquid at $\Delta T/T_m = 0.20$,  including all the atoms from 50 configurations for each MD trajectory. The reference histograms in the $\bar{q}_4-\bar{q}_6$ map show little overlap and we can assign a crystal structure to each particle.  To this end, the structure with the largest probability is assigned to the corresponding $\bar{q}_4-\bar{q}_6$ values of a particle. If all crystal probabilities are smaller than $10^{-5}$ and the liquid probability vanishes, 
%but the  solidity criterion of ten Wolde {\it et al.} is fulfilled, 
a particle is labeled  as pre-structured liquid particles. The pre-structured liquid particles show less symmetry than the crystal structures but higher bond-orientational order than the liquid, thus their $\bar{q}_6$ and $\bar{q}_4$ values lie in between the liquid and crystal regions.~\cite{DLeines2017}. In this work, all the pre-structured particles included in our analysis are selected to have more than 50\% pre-structured liquid neighbours, in order to reduce solid-liquid interface effects in the structural analysis.

%\clearpage
%\section{ Minimum distance distributions and $\hat{q}_4-\hat{q}_6$ values for pre-ordered clusters  in the liquid}
%
%{\bf $\hat{q}_4-\hat{q}_6$ values for pre-ordered atoms in the liquid}
\section{Additional graphs}
\begin{figure*}[htb]
\centering
\includegraphics[width=10.2cm,clip=true]{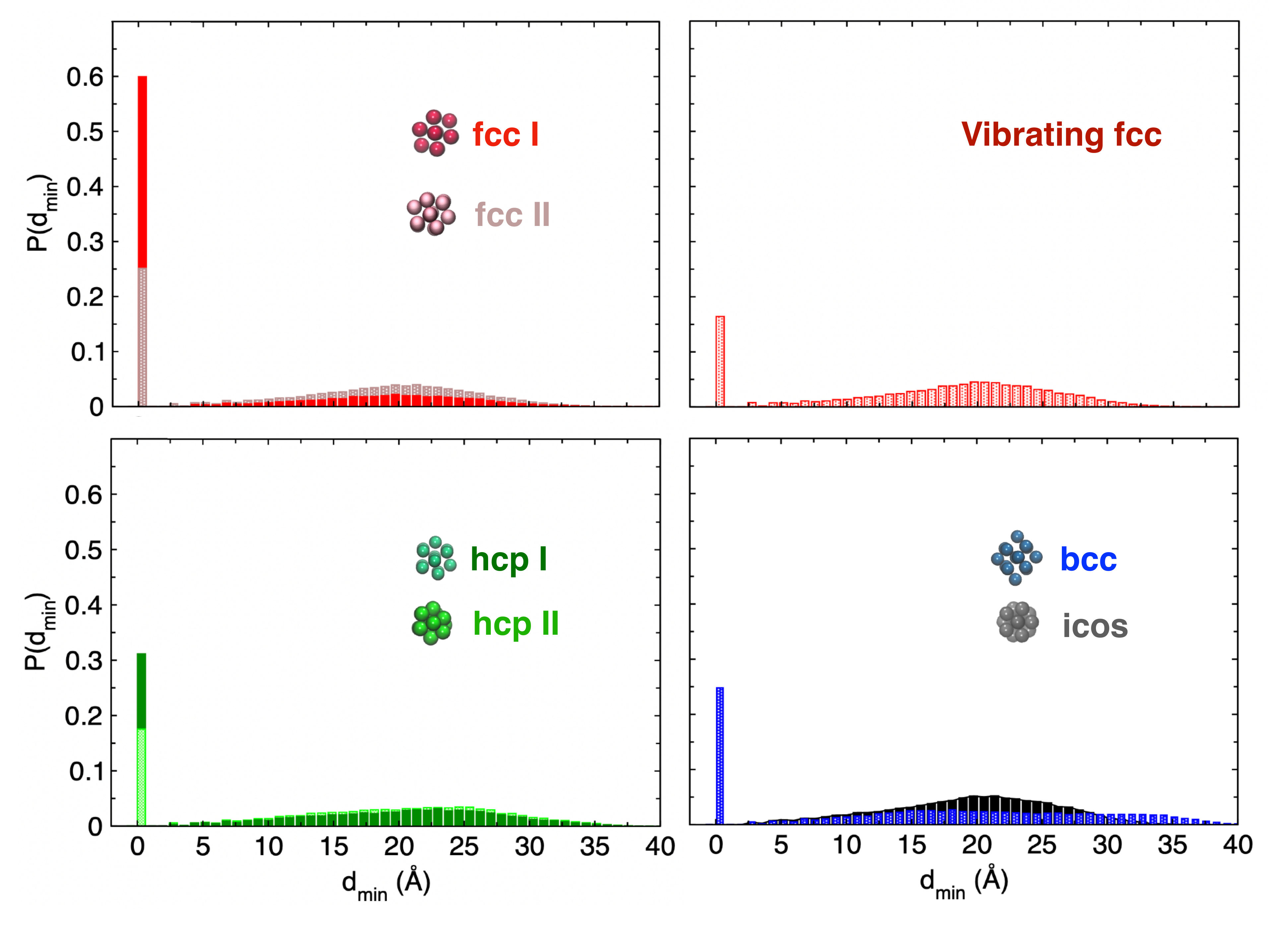} \caption{\label{fig1supp} Histograms of the minimum distance from the seed to the largest pre-critical cluster in liquid Ni. The distributions are computed from five independent MD simulations of liquid, 5 ns each. The distributions show good agreement with the histograms of the number of seed atoms included in the largest cluster, i.e. for seeds with higher crystallinity, like fcc I,  fcc II and hcp I, the seeds are more frequently part of the clusters ($d_{\text{min}}=0$), while for other seeds less commensurable and with lower cristallinity, the clusters emerge frequently far from the seeds ($d_{\text{min}} \geq 20.0$~\AA)
}
\end{figure*}

\begin{figure*}[htb]
\centering
\includegraphics[width=14.5cm,clip=true]{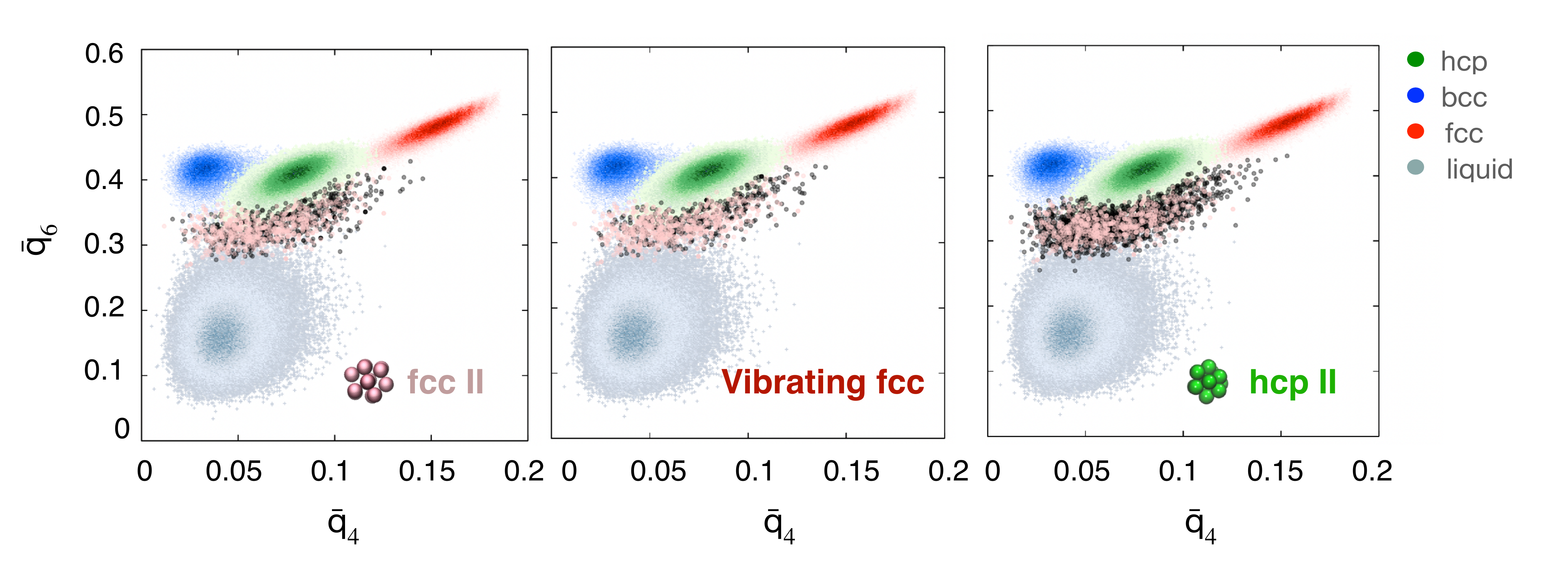}
\caption{\label{fig2supp}   $\bar{q}_4-\bar{q}_6$ values for  pre-ordered liquid particles that belong to the largest pre-critical cluster that emerges within homogeneous liquid (scattered pink dots) and at fcc II, vibrating fcc and hcp II seeds (scattered black dots) in the liquid. 
}
\end{figure*}
%\section{Structural composition of the growing nucleus}

\begin{figure*}[tb]
\centering
\includegraphics[width=15.2cm,clip=true]{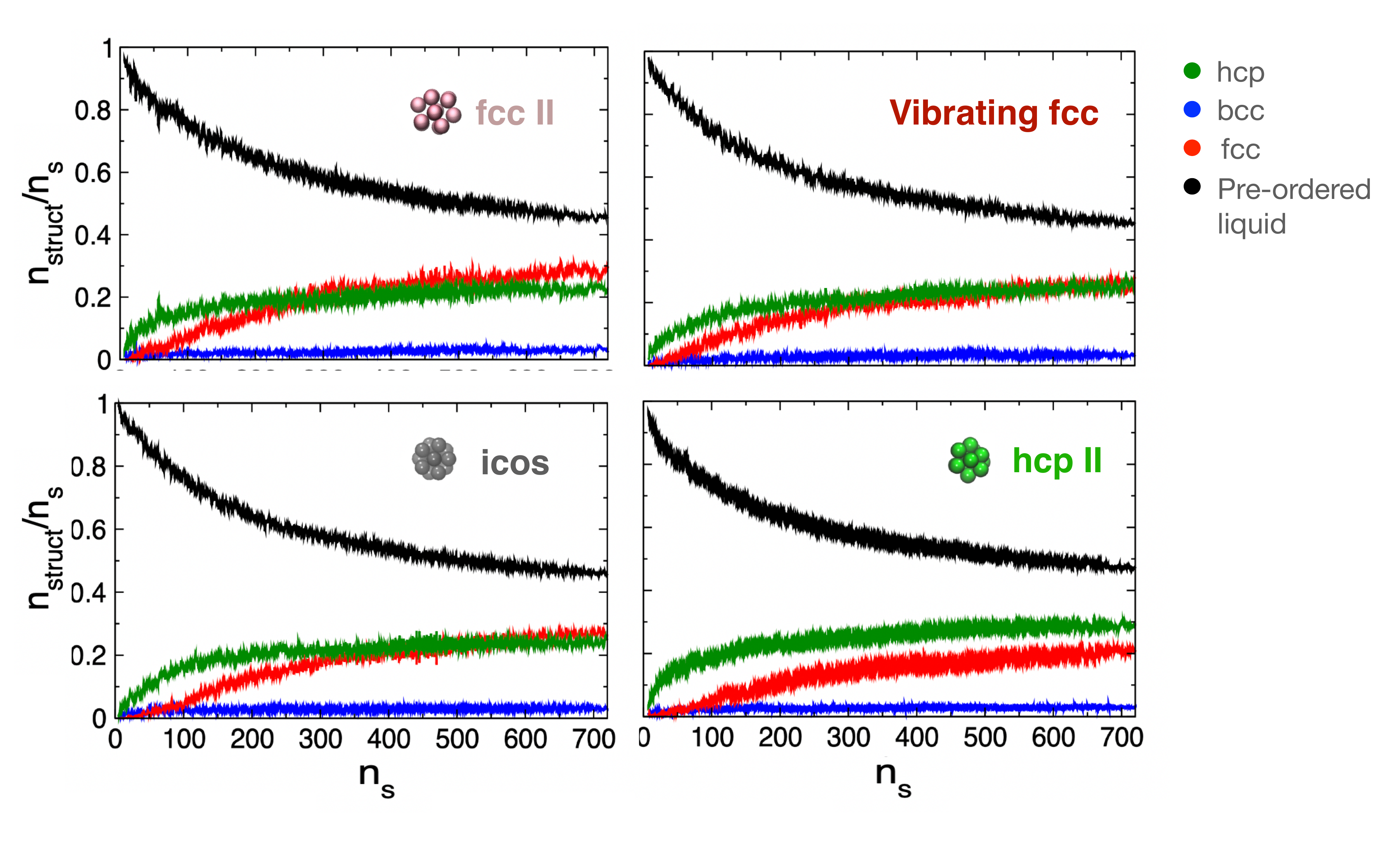}
\caption{\label{fig3supp} Average structural composition of the growing nucleus. The fractions of bcc (blue), hcp (green), fcc (red), and pre-structured liquid (black) particles in the largest growing solid cluster at 20\% undercooling are shown. The average fractions calculated from all configurations of at least 300 liquid-solid trajectories in the TPE.
}
\end{figure*}

\begin{figure*}[tb]
\centering
\includegraphics[width=16.2cm,clip=true]{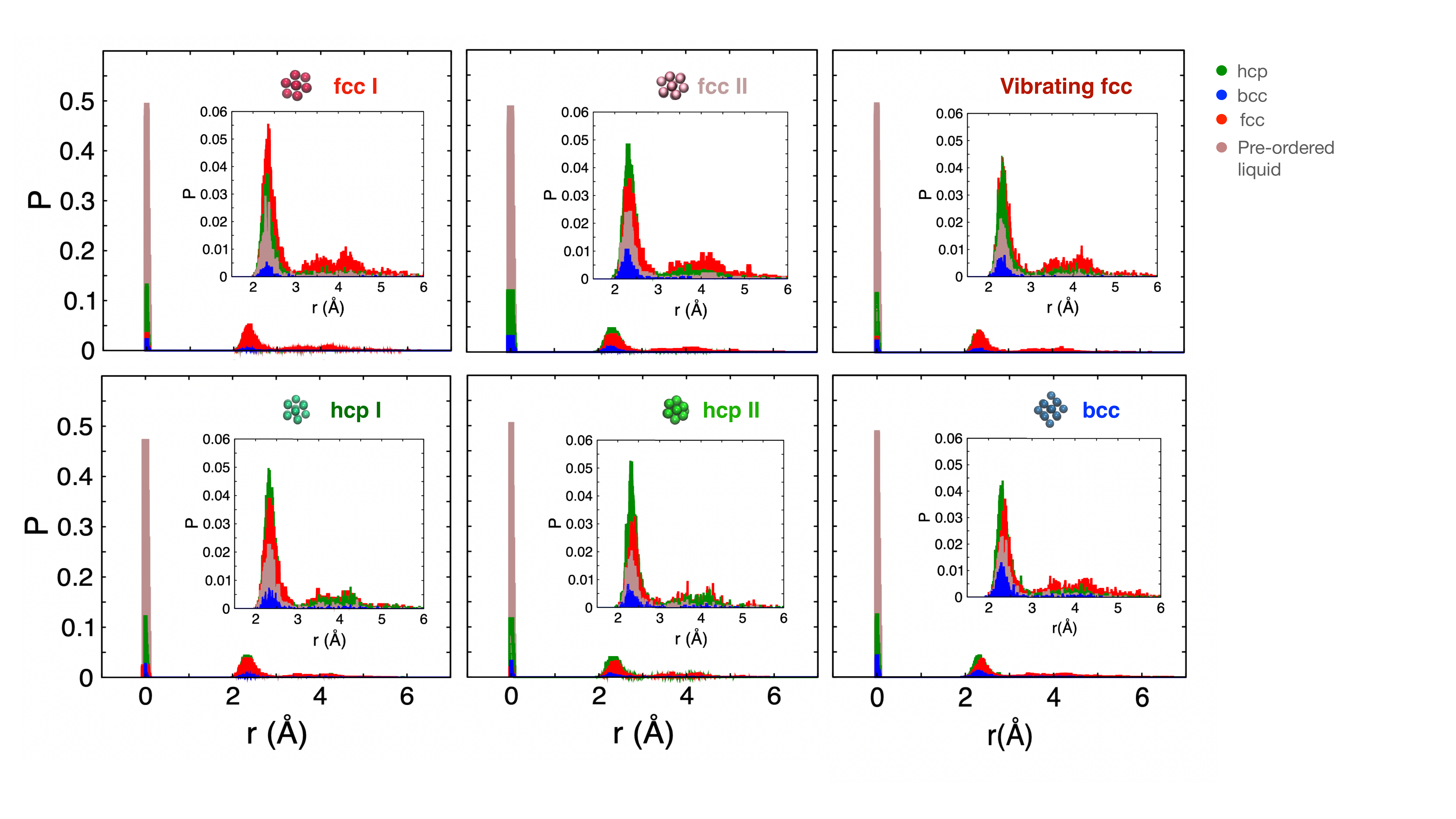}
\caption{\label{fig4supp} Structural composition distributions for critical clusters obtained from the transition state ensemble, as a function of the minimum distance from the cluster surface $r$, in the presence of different seeds (fcc I, fcc II, vibrating fcc seed, bcc, hcp I and hcp II) at 20\% undercooling. The distance $r$ is defined as the minimum distance of an atom in the largest solid cluster to a surface particle, identified as a solid particle in the largest cluster that has at least one liquid neighbour.  The distributions of fcc (red), bcc (blue), hcp (green), and pre-structured (light brown) particles in the largest cluster are shown for the surface ($r=0.0$) and the core of the nuclei (zoomed-in inset shows the core structure of the clusters). 
A clear difference in the composition of the core is observed for the different seeds; in particular, fcc seeds favour the formation of fcc in the core, whereas hcp seeds lead to more hcp in the core of the critical clusters.
Distributions were obtained from at least 500 transition state configurations.  
}
\end{figure*}

\begin{figure*}[tb]
\centering
\includegraphics[width=14.2cm,clip=true]{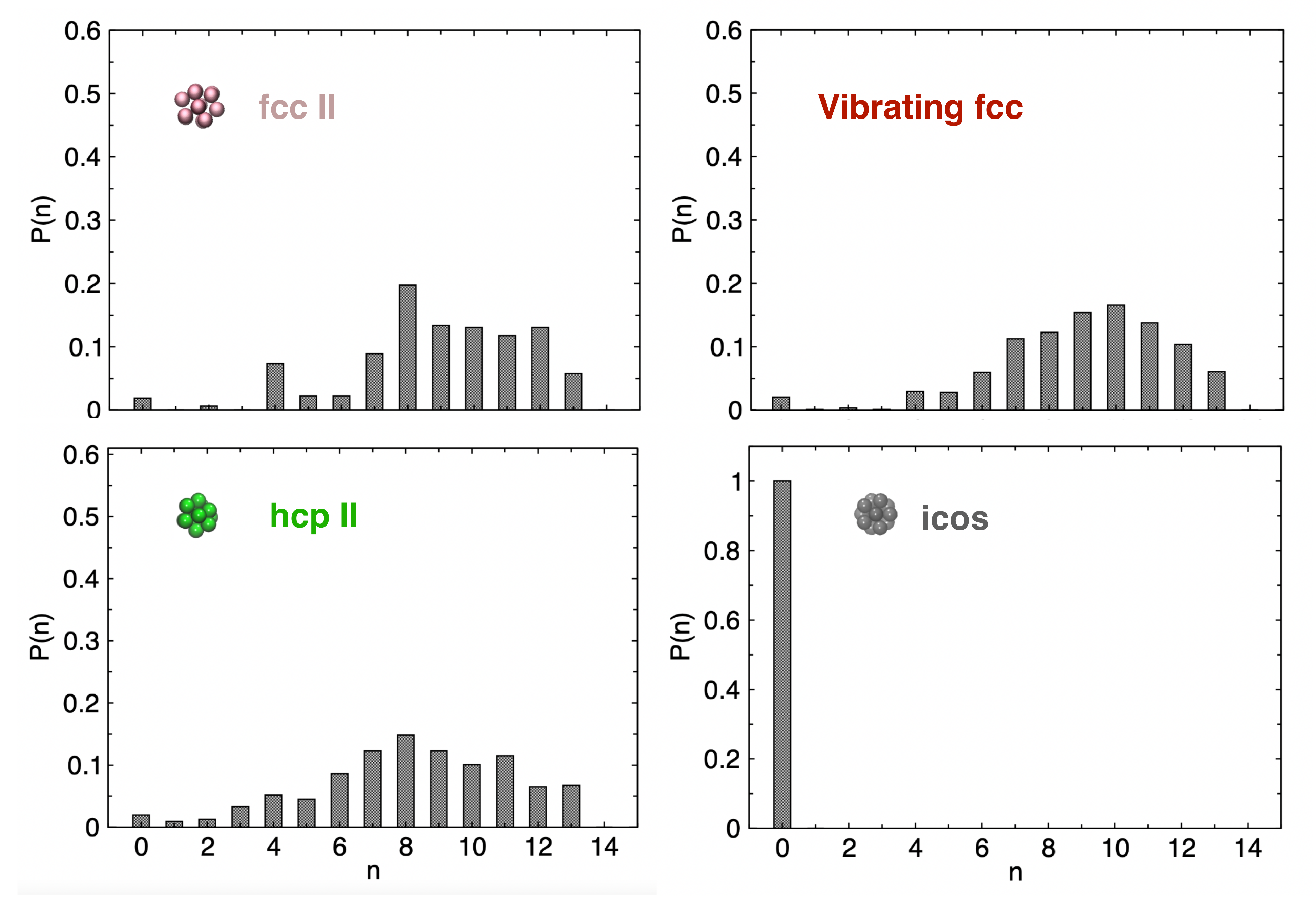}  
\caption{\label{fig2supp} Distribution of number of seed atoms included in the pre-critical clusters ($n_s$=50) that successfully nucleate and crystallize (effective precursors) in the presence of different seeds. At least 500 pre-critical clusters configurations were included in the calculation of the distributions and obtained from the transition path ensemble.
}
\end{figure*}

%\subsection{Distributions of number of seed atoms and $\hat{q}_4,\hat{q}_6$ values in effective precursors of the TPE }

\begin{figure*}[tb]
\centering
\includegraphics[width=14.2cm,clip=true]{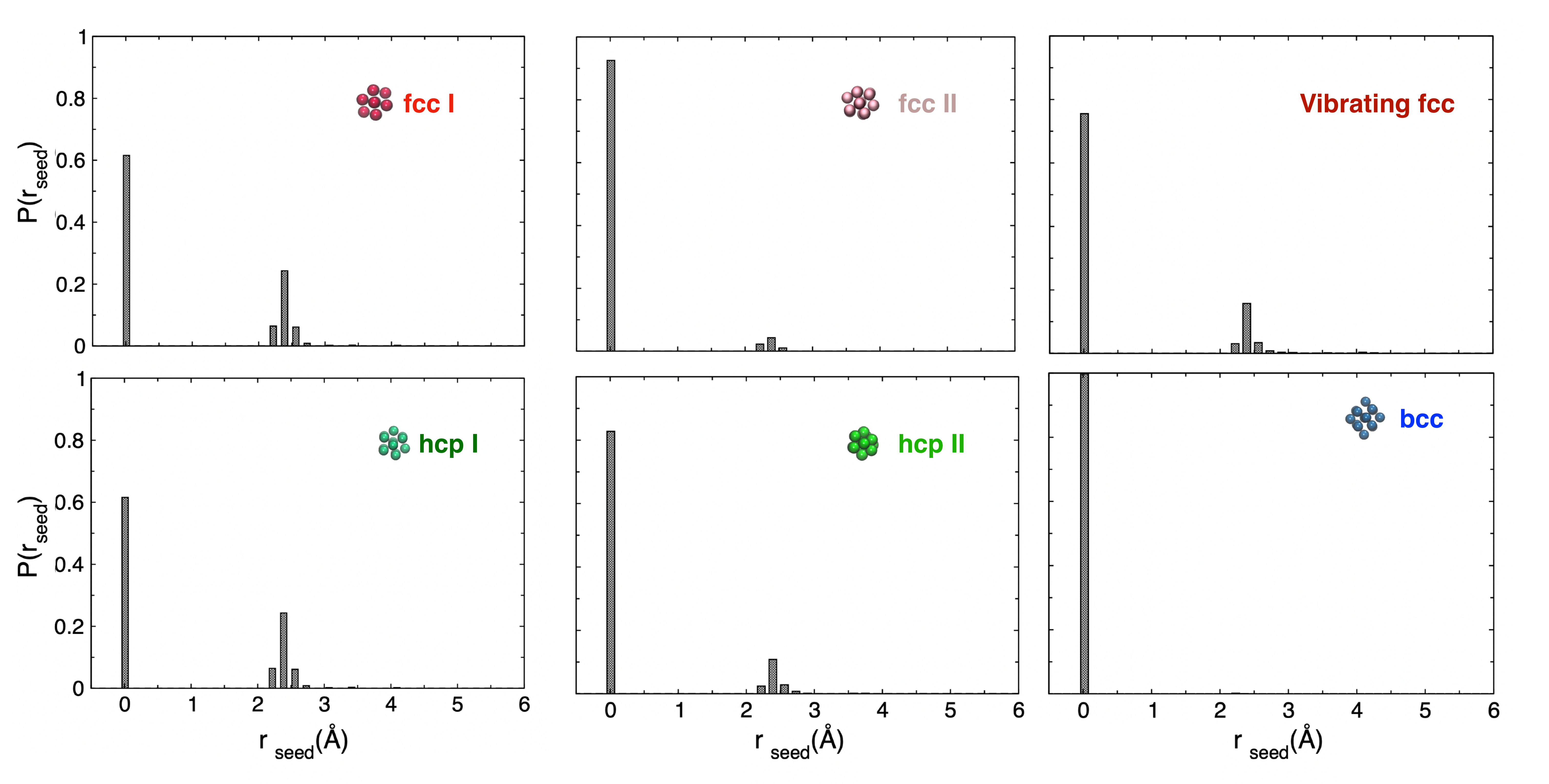}  
\caption{\label{fig5supp} Spatial location of a seed in the critical solid clusters. The plots show the distribution of the minimum distance from the seed to the surface atoms in the critical nucleus.  The distance $r$ is the minimum distance of a seed atom in the critical solid cluster to a surface particle (a solid particle in the critical cluster that has at least one liquid neighbour). Seeds are predominantly located at the surface with $r=0$.  The distributions include at least 500 configurations obtained from the transition state ensembles, that include at least one atom of the seed in the solid cluster.
}
\end{figure*}

\begin{figure*}[tb]
\centering
\includegraphics[width=14.2cm,clip=true]{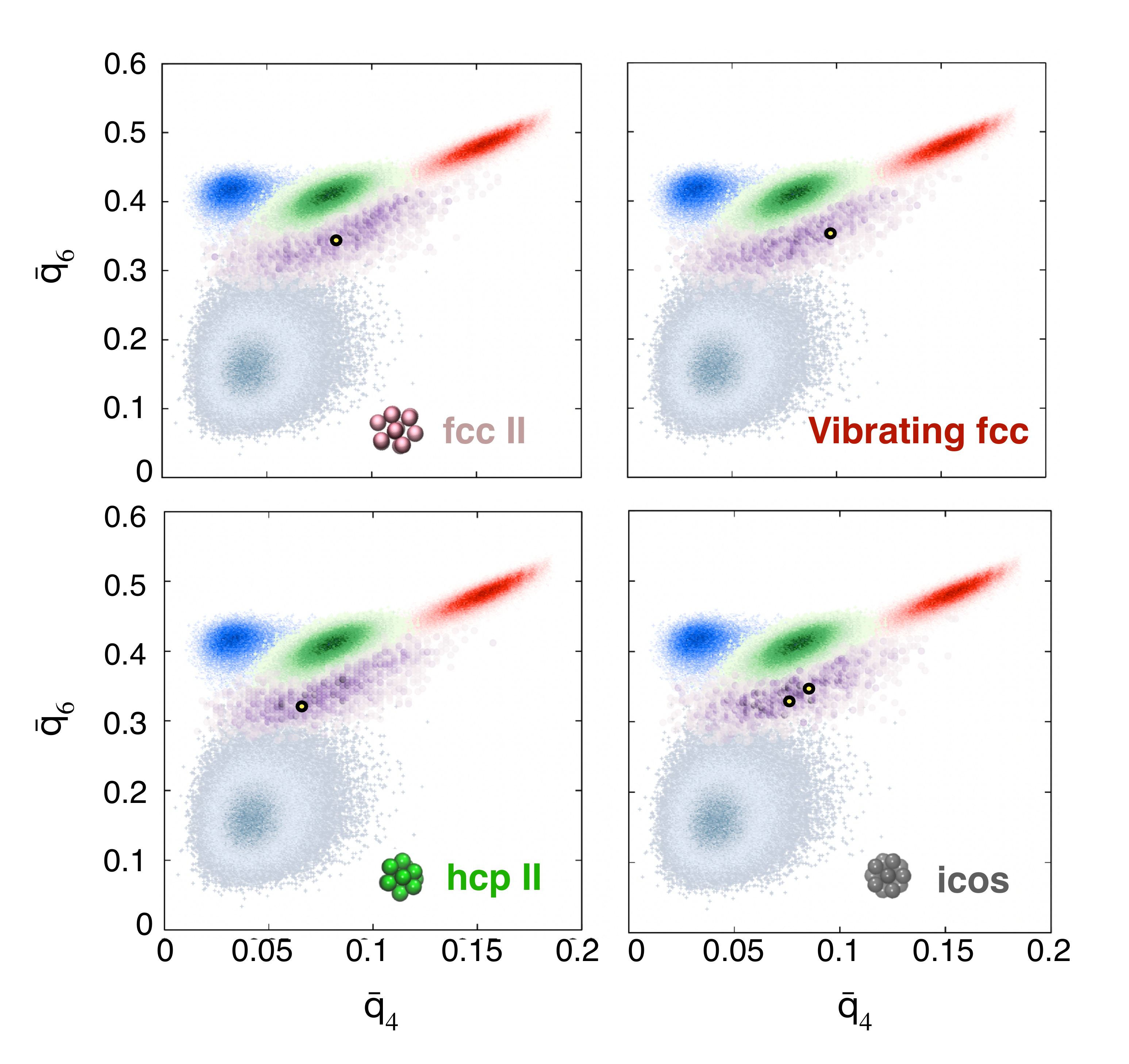}  
\caption{\label{fig6supp} $\bar{q}_4,\bar{q_6}$ distributions of the effective precursors obtained from at least 500 pre-critical clusters of the TPE, in the presence of different seeds. The yellow circle indicates the maximum of the distribution.
}
\end{figure*}

\begin{figure*}[tb]
\centering
\includegraphics[width=12.2cm,clip=true]{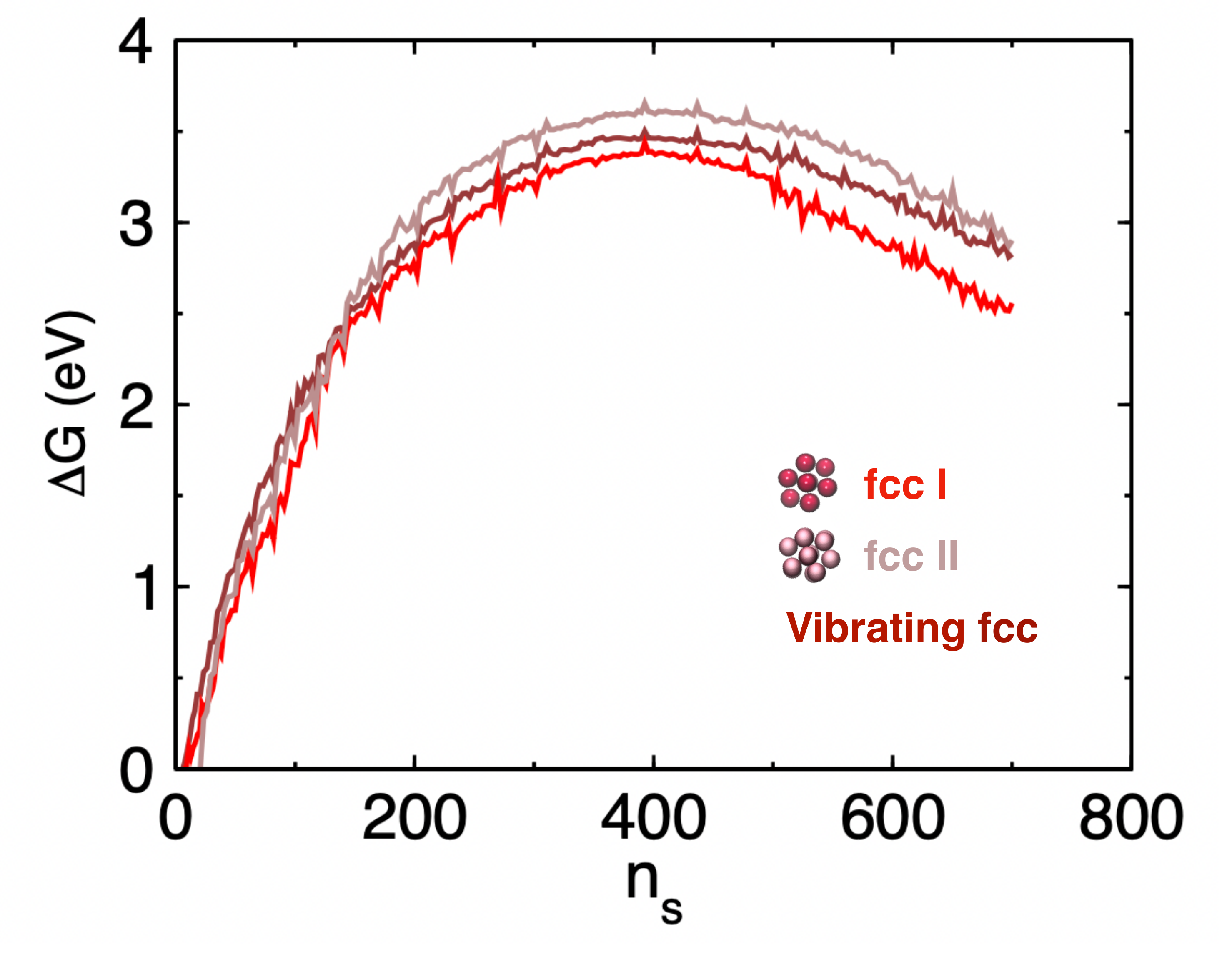}  
\caption{\label{fig7supp} Free energy profiles and nucleation barriers in the presence of various fcc seeds. The profiles show that even if the seeds share a common crystal structure, their variations in crystallinity result in significant differences in the nucleation barriers. In agreement with the observed template precursor mechanism of heterogeneous nucleation in Ni, the fcc seed with higher crystalllinity (fcc~I) that enhances the formation of effective precursors the most  reduces the barrier more than  fcc seeds with lower crystallinity (fcc~II and vibration fcc seed). Moreover, as the vibration of the fcc seed allows for seed crystallinity values both higher and lower than fcc~II, it is expected that the vibrating fcc seed is relatively more efficient than fcc II to enhance the nucleation probability, but less efficient than fcc I, as corroborated by the nucleation barriers in this plot. 
}
\end{figure*}

% \begin{figure}[htbp]
 % \centering
     %\centering
%     \includegraphics[width=1.0\linewidth]{images/voronoi_fs.png}
 %    \caption{Comparison of polyhedra in (a) liquid, (b) pre-structured, and (c) solid particles for Finnis-Sinclair potential at 20\% undercooling.
 %    }
 %    \label{fig:poly_fs}
% \end{figure}
 
\section{Movie}

Movie of representative nucleation trajectory in the presence of the fcc~II seed, obtained from the path ensemble. The pre-structured liquid region (blue transparent surface) forms near the seed, preceding the emergence of the crystalline clusters  within the center of the precursors and composed of fcc (red atoms) and random-hcp (green atoms). The seed fcc II (blue atoms) is mostly located at the surface of the cluster and surrounded predominantly by pre-structured liquid and random-hcp.

%\bibliographystyle{unsrt}
%\bibliography{references}